\documentclass{emulateapj}
\usepackage{graphicx}
\usepackage{wasysym}
\usepackage{natbib}
\usepackage[colorlinks,urlcolor=cyan,citecolor=blue,linkcolor=blue]{hyperref} 

\slugcomment{Accepted to ApJ}
\shortauthors{} \shorttitle{{\it Spitzer} Observations of The Low-density Exo-Neptune GJ\,3470\,b}
\begin{document}

\title{{\it Spitzer} Observations of GJ\,3470\,b: a Very Low-density Neptune-size Planet \\ Orbiting a Metal-rich M dwarf$^{\star}$}

\author{Brice-Olivier~Demory\altaffilmark{1}, Guillermo~Torres\altaffilmark{2}, Vasco~Neves\altaffilmark{3,4,5},  Leslie~Rogers\altaffilmark{6}, Micha{\"e}l~Gillon\altaffilmark{7}, Elliott~Horch\altaffilmark{8,9}, Peter~Sullivan\altaffilmark{10}, Xavier~Bonfils\altaffilmark{5}, Xavier~Delfosse\altaffilmark{5}, Thierry~Forveille\altaffilmark{5},  Christophe~Lovis\altaffilmark{11}, Michel~Mayor\altaffilmark{11}, Nuno~Santos\altaffilmark{3,4}, Sara~Seager\altaffilmark{1,10}, Barry~Smalley\altaffilmark{12}, Stephane~Udry\altaffilmark{11}}

\altaffiltext{1}{Department of Earth, Atmospheric and Planetary Sciences, Massachusetts Institute of Technology, 77 Massachusetts Ave., Cambridge, MA 02139, USA. demory@mit.edu}
\altaffiltext{2}{Harvard-Smithsonian Center for Astrophysics, 60 Garden St., Cambridge MA 02138, USA}
\altaffiltext{3}{Centro de Astrof\'{\i}sica, Universidade do Porto, Rua das Estrelas, 4150-762 Porto, Portugal}
\altaffiltext{4}{Departamento de F\'{\i}sica e Astronomia, Faculdade de Ci\^encias, Universidade do Porto, Rua do Campo Alegre, 4169-007 Porto, Portugal}
\altaffiltext{5}{UJF-Grenoble 1 / CNRS-INSU, Institut de Plan\'etologie et d'Astrophysique de Grenoble (IPAG) UMR 5274, Grenoble, F-38041, France.}
\altaffiltext{6}{Department of Astrophysics, California Institute of Technology, MC 249-17, Pasadena, CA 91125, USA}
\altaffiltext{7}{Institut d'Astrophysique et de G\'eophysique, Universit\'e de Li\`ege, All\'ee du 6 Ao\^ut, 17, Bat. B5C, Li\`ege 1, Belgium.}
\altaffiltext{8}{Department of Physics, 501 Crescent Street, Southern Connecticut State University, New Haven, CT 06515, USA}
\altaffiltext{9}{Visiting Astronomer, Kitt Peak National Observatory, National Optical Astronomy Observatories, which is operated by the Association of Universities for Research in Astronomy, Inc. (AURA) under cooperative agreement with the National Science Foundation.}
\altaffiltext{10}{Department of Physics and Kavli Institute for Astrophysics and Space Research, MIT, 77 Massachusetts Avenue, Cambridge, MA 02138, USA.}
\altaffiltext{11}{Observatoire de Gen\`eve, Universit\'e de Gen\`eve, 51 ch. des Maillettes, CH-1290 Versoix, Switzerland}
\altaffiltext{12}{Astrophysics Group, Keele University, Staffordshire, ST55BG, UK}
\altaffiltext{$\star$}{This paper includes data gathered with the 6.5 meter Magellan Telescopes located at Las Campanas Observatory, Chile.}

\begin{abstract}

We present {\it Spitzer}/IRAC 4.5\,$\mu$m transit photometry of GJ\,3470\,b, a Neptune-size planet orbiting a M1.5 dwarf star with a 3.3-day period recently discovered in the course of the HARPS M-dwarf survey. We refine the stellar parameters by employing purely empirical mass-luminosity and surface brightness relations constrained by our updated value for the mean stellar density, and additional information from new near-infrared spectroscopic observations. We derive a stellar mass of $M_{\star} =  0.539^{+0.047}_{-0.043} \,M_{\odot}$ and a radius of $R_{\star} = 0.568^{+0.037}_{-0.031}\,R_{\odot}$. We determine the host star of GJ\,3470\,b to be metal-rich, with a metallicity of ${\rm [Fe/H]} = +0.20 \pm 0.10$ and an effective temperature of $T_{\rm eff} = 3600 \pm100$\,K.
The revised stellar parameters yield a planetary radius $R_{p} = 4.83_{-0.21}^{+0.22}\,R_{\oplus}$ that is 13\% larger than the value previously reported in the literature. We find a planetary mass $M_{p} = 13.9^{+1.5}_{-1.4}\,M_{\oplus}$ that translates to a very low planetary density, $\rho_{p}= 0.72^{+0.13}_{-0.12}$ g\,cm$^{-3}$, which is 33\% smaller than the original value. With a mean density half of that of GJ\,436\,b, GJ\,3470\,b is an example of a very low-density low-mass planet, similar to Kepler-11\,d, Kepler-11\,e, and Kepler-18\,c but orbiting a much brighter nearby star that is more conducive to follow-up studies.

\end{abstract}


\section{Introduction}

In the regime of low mass exoplanets only a handful of those known to periodically pass in front of their host stars have transits that are deep enough and orbit parent stars that are bright enough to make them amenable to extensive follow-up observations. The \textit{Kepler} mission has recently announced a harvest of more than 2,700 planetary candidates identified since the launch of the spacecraft in 2009 \citep{Batalha:2012}. About ten percent are Jupiter-size planets with radii between 0.7 and 2.0 Jupiter radii, while more than 55\% are Neptune-size planets with radii between 2 and 6 Earth radii. On the other hand, among the 241 confirmed transiting exoplanets (coming mainly from ground-based surveys), 62\% are Jupiter-size planets with radii between 0.7 and 2.0 Jupiter radii\footnote{Source: \url{http://www.exoplanets.org}}. It is now clear from \textit{Kepler} and other studies that short-period Jupiter-size objects make up a relatively small fraction of the exoplanet population \citep[e.g.,][]{Howard:2010,Wittenmyer:2011,Howard:2012} . This stark contrast between confirmed exoplanets and the large underlying population glimpsed by \textit{Kepler} has motivated intense efforts towards the characterization of smaller planets, in order to reach a comparable state of knowledge to what has been learned about the hot-Jupiter population. These efforts began already several years ago with the launch of a number of ground-based projects dedicated to M-dwarf monitoring using both spectroscopy \citep[e.g., the HARPS program;][]{Bonfils:2013} and photometry \citep[e.g., MEarth;][]{Nutzman:2008}. Planets orbiting M-dwarf stars offer the possibility to probe smaller planets for a given transit depth, because of the favorable star-to-planet radius ratio. GJ\,436\,b \citep{Butler:2004,Gillon:2007c} and GJ\,1214\,b \citep{Charbonneau:2009} are the smallest planets orbiting M stars with $K$ magnitude brighter than nine, enabling detailed follow-up studies both from the ground and from space \citep[e.g.,][]{Stevenson:2010,Bean:2010}.

In the Neptune-mass range, thanks to its relatively large transit depth and host star brightness, GJ\,436\,b remains a ``Rosetta stone'' for our understanding of a whole class of exoplanets, shown to be ubiquitous in our Galaxy. With a mass 22 times that of the Earth and a radius 4 times larger than our home planet, GJ\,436\,b has a relatively high density \citep[$\rho_{p} = 1.69^{+0.14}_{-0.12}$ g\,cm$^{-3}$;][]{Torres:2008}, suggesting the presence of a massive core made of silicates and/or ices. However, a H/He envelope is needed to reproduce its observed radius \citep[e.g.,][]{Figueira:2009,Rogers:2010}. The improvement in the planetary radius of this object brought about by \textit{Spitzer} observations placed significant constraints on the range of possible compositions of GJ\,436\,b's interior. A key question that still needs to be addressed, however, is the extent to which GJ\,436\,b is representative of the entire exo-Neptune population.

GJ\,3470\,b is a new transiting Neptune-size planet discovered in the past year \citep{Bonfils:2012a}. It orbits a $K_s = 7.99$ mag, M1.5 dwarf with a period of 3.337 days. With a published mass of $14.0 \pm 1.7$ Earth masses and a radius of $4.2 \pm 0.6$ Earth radii \citep{Bonfils:2012a}, GJ\,3470\,b has a mean density $\rho_{p}= 1.07\pm0.43$ g\,cm$^{-3}$ that is significantly smaller than that of GJ\,436\,b. The \textit{Kepler} mission confirmed several of these so-called ``low-density Neptunes''. The first two were Kepler-11\,d and e \citep{Lissauer:2011}, both belonging to the most populated transiting planet system known to date, and the third was Kepler-18\,c \citep{Cochran:2011a}, also a member of a multi-planet system. These objects represent the tip of the iceberg, as several hundred Neptune-size planet candidates have already been detected by \textit{Kepler} and await confirmation. Unfortunately, most of these \textit{Kepler} planets orbit faint stars and exhibit shallow transit depths that render follow-up studies very challenging, if not impractical altogether. Aside from {\it Kepler}, the ground-based survey HAT discovered the low-density Neptune HAT-P-26\,b \citep{Hartman:2011} which, until the discovery of GJ\,3470\,b, represented the most promising target for follow-up studies. However, as compared to GJ\,3470\,b, the smaller planet-to-star area ratio coupled with the lower brightness of its larger K1 host star ($K=9.6$) makes HAT-P-26\,b a less favorable target for follow-up studies. GJ\,3470\,b therefore presents an ideal opportunity to investigate the internal structure, atmospheric composition, and possible formation pathways of low-density Neptune-size planets \citep[e.g.,][]{Rogers:2011}.

All transit photometry available so far for GJ\,3470 has been collected from the ground. While these time-series confirm the transiting nature of GJ\,3470\,b, they do not precisely constrain the transit parameters, resulting in poorly determined planetary properties. 
We present in this paper the analysis of two transits of GJ\,3470\,b obtained with the {\it Spitzer} Space Telescope at 4.5 $\mu$m in the frame of our DDT program \citep{Demory:2012a}, submitted shortly after GJ\,3470\,b's discovery. These data yield a significant refinement of the system parameters. The paper is organized as follows. Section~\ref{obs} describes the observations and data reduction, while Section~\ref{analysis} presents the photometric and spectroscopic data analyses. Section~\ref{stellar} is dedicated to the stellar characterization, and the resulting planetary parameters are reported in Section~\ref{planet}. We discuss GJ\,3470\,b's internal structure and composition in Section~\ref{internal}.

\section{Observations and Data Reduction}
\label{obs}

\subsection{Spitzer IRAC 4.5-$\mu$m photometry}

\begin{figure}
\epsscale{1.0}
\plotone{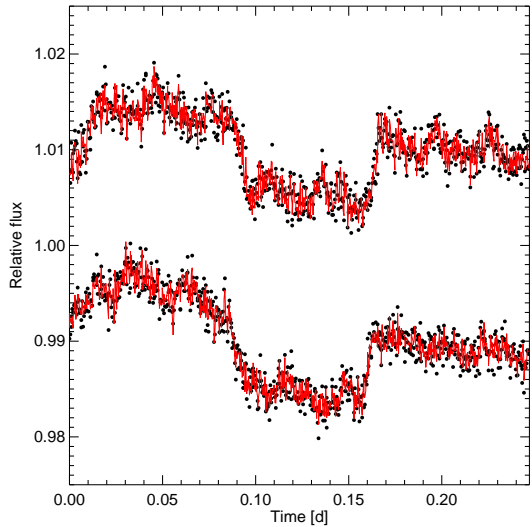}
\figcaption[]{Spitzer/IRAC 4.5$\mu$m photometry. Raw photometry from the two AORs is displayed after normalization. The optimal baseline model (logarithmic ramp model added to a second-order polynomial fit for the centroid position and a time-dependent trend (see Section~\ref{analysis}) is superimposed for each AOR in red.\label{fig:phot1}}
\end{figure}

We observed two consecutive transits of GJ\,3470\,b at 4.5 $\mu$m using {\it Spitzer}'s InfraRed Array Camera \cite[IRAC;][]{Fazio:2004a}. Observations took place on 11 June 2012 and 15 June 2012 UTC as part of our DDT program PID 80261. For each transit we obtained 780 sets of 64 subarray frames each, with an exposure time of 0.40\,s per frame. Each Astronomical Observation Request (AOR) lasted 6.5 hours, including 30 minutes overhead for the Pointing Calibration and Reference Sensor peak-up sequence. This step allowed GJ\,3470 to be precisely slewed on the position of maximum sensitivity on the 4.5$\mu$m channel subarray field-of-view  \citep{Ingalls:2012,Demory:2012}. All data were processed by the {\it Spitzer} pipeline version S19.1.0, which produced the basic calibrated data  necessary for our reductions. We first convert fluxes from the {\it Spitzer} units of specific intensity (MJy/sr) to photon counts, and transform the data timestamps from BJD$_{\rm UTC}$ to BJD$_{\rm TDB}$ following \citet{Eastman:2010a}. We then perform aperture photometry on each subarray image using the \textsc{APER} routine from the IDL Astronomy User's Library\footnote{\url{http://idlastro.gsfc.nasa.gov/contents.html}}. We compute the stellar fluxes in aperture radii ranging between 1.8 and 4.0 pixels, the best results being obtained with an aperture radius of 3 pixels. We use background annuli extending from 11 to 15.5 pixels from the Point Response Function center. For each block of 64 subarray images, we discard the discrepant values for the measurements of flux, background, and $x$-$y$ centroid positions using a 10-$\sigma$ median clipping for the four parameters. We then average the resulting values, the photometric errors being taken as the uncertainties on the average flux measurements. At this stage, a 50-$\sigma$ clipping moving average is used on the resulting light curve to discard obviously discrepant subarray-averaged fluxes. Close examination of the resulting time-series reveals a sharp increase of the background and stellar fluxes, corresponding to the well-known ``ramp'' effect seen in other warm {\it Spitzer} observations \citep[see, e.g.,][]{Knutson:2012}. The raw photometry for both AORs is shown on Figure~\ref{fig:phot1}.

\subsection{WIYN Speckle Observations}

We supplemented our GJ\,3470\,b {\it Spitzer} photometry with speckle observations to explore the possibility of blended companions at close angular separations from GJ\,3470. Speckle observations of GJ\,3470 were obtained at the WIYN 3.5-m telescope on 2 December 2012. The camera used was the Differential Speckle Survey Instrument, which is described by \citet{Horch:2009}. It is a dual-channel instrument that records images in two colors simultaneously. In the case of this observation, the filters used had center wavelengths of 692\,nm and 880\,nm, with filter widths of 40 and 50\,nm, respectively. A speckle sequence of 3000 50-ms frames was taken on the target, followed by 1000 frames taken on a bright point source  (HR\,3163) located near in the sky to GJ\,3470. These latter data are used as an estimate of the speckle transfer function for deconvolution in the reduction process. Reconstructed images are formed from the speckle data using the technique of bispectral analysis, which is described, e.g., by \citet{Horch:2012}. We then analyze the final images to determine the detection limits of faint companions near GJ\,3470 using the technique described in the same paper.

\begin{figure}
\epsscale{1.0}
\plotone{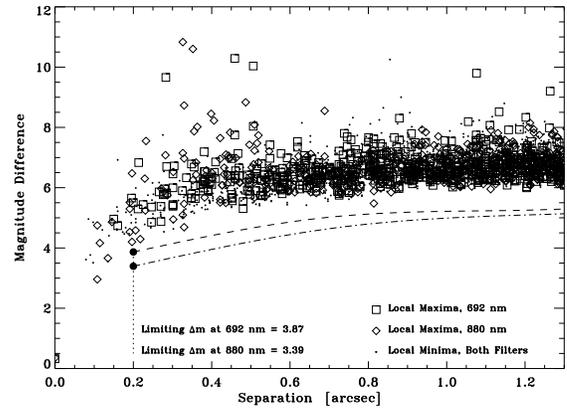}
\figcaption[]{Detection limit analysis of GJ 3470 based on speckle data. The squares and diamonds indicate the magnitude differences of local maxima in each reconstructed image as a function of separation from the central star, and points are local minima. The curves indicate the 5-$\sigma$ detection limit based on the statistics of the these maxima and minima. The dashed curve is the result for the 692\,nm reconstructed image, and the dot-dash curve is the result for the 880\,nm image. These curves indicate a greater than 4-magnitude sensitivity in the limiting $\Delta m$ at most separations. \label{figspeck}}
\end{figure}

Figure~\ref{figspeck} shows these detection limits based on the final diffraction-limited images in each filter. It is clear that there is no companion to the limit of our detection capabilities at a separation greater than 0\farcs2. At 0\farcs2 the limiting $\Delta m$ for the 692\,nm image is 3.87 mag, and for the 880\,nm image it is 3.39 mag. Inside of this limit, as one approaches the central star, the limiting $\Delta m$ becomes smaller as the peaks and valleys of the reconstructed image get larger. In studying the two images, we find that none of the peaks near the central star are in the same position in both images, which is a good indication that they are probably not real stars but noise peaks. One of the advantages of the two independent channels in the instrument is to see if the positions of faint peaks match. We conclude that, to the limit of our detection at WIYN, there is no resolvable companion.

\subsection{Magellan/FIRE Near-infrared Spectroscopy}
\label{fire}

Spectroscopic properties such as the effective temperature, $T_{\rm eff}$, and metallicity, [Fe/H], needed to establish the physical parameters of the parent stars of transiting planets have usually been difficult to determine for M dwarfs. Several studies in the past year have presented calibrations of [Fe/H] or $T_{\rm eff}$ in terms of easily measurable spectroscopic indices in the $H$-band and $K$-band regions that represent a significant advancement in the field. Toward this goal, we obtained a near-infrared spectrum of GJ\,3470 on 12 November 2012 with the Folded-port Infrared Echellette (FIRE) spectrograph at the 6.5-m Magellan Baade telescope. FIRE delivers $R=6000$ spectra from 0.83 to 2.5\,$\mu$m in a single-object, cross-dispersed setup \citep{Simcoe:2008}. We used an exposure time of 8.5 minutes on GJ\,3470 and 105 seconds on the A0V telluric standard HD\,58296. We reduce the data using FIRE's pipeline FIREHOSE, which employs the methods of \citet{Vacca:2003} for telluric correction. An internal ThAr lamp provides wavelength calibration of both the GJ\,3470 and HD\,58296 spectra. 
The signal-to-noise ratio in the reduced spectrum is $>200$ in the $K$ band, where H$_{2}$O features near 2.21 and 2.26\,$\mu$m fall on order 20 of FIRE's 21 cross-dispersed grating orders. Strong OH emission lines from the sky (which can introduce shot noise and residuals from sky subtraction) do not fall directly on these features. The measurement of various spectroscopic indices from this FIRE observation is described in Sect.~\ref{fire_analysis}.

\section{Data Analysis}
\label{analysis}

\subsection{{\it Spitzer} Photometry}

\subsubsection{Baseline Model Selection}

We first perform an individual analysis of each {\it Spitzer} AOR to determine the optimal baseline model, which accounts for time- and position-dependent systematic effects relevant to our IRAC 4.5\,$\mu$m observations. We employ for this purpose our adaptive Markov-Chain Monte Carlo (MCMC) implementation described by \citet{Gillon:2010a}. We test six baseline models of increasing complexity, and compare their Bayesian information criteria \citep[BIC; see, e.g.,][]{Gelman:2003} to choose the baseline model that yields the highest marginal likelihood. We correct for the well-known ``pixel-phase'' effect using a second- to fourth-order $x$-$y$ position-dependent polynomial, while the ``ramp'' is corrected using a second-order logarithmic model. We also check for time-dependent trends of instrumental and/or stellar origin by adding linear or quadratic functions of time to our baseline models. We additionally explore the correlation of the stellar flux and background time-series with the full width at half maximum (FWHM) of the point response function \citep{Demory:2012}. We find for both AORs the lowest BIC to correspond to a model including a second-order position-dependent polynomial, a second-order logarithmic ramp, and a time-dependent linear trend. Our analysis yields an RMS of 362\,ppm and 369\,ppm per 5 min interval in the first and second AOR, respectively, with negligible contribution from correlated noise.

\subsubsection{Determination of the Stellar Density}
\label{density}

\begin{figure}
\epsscale{1.0}
\plotone{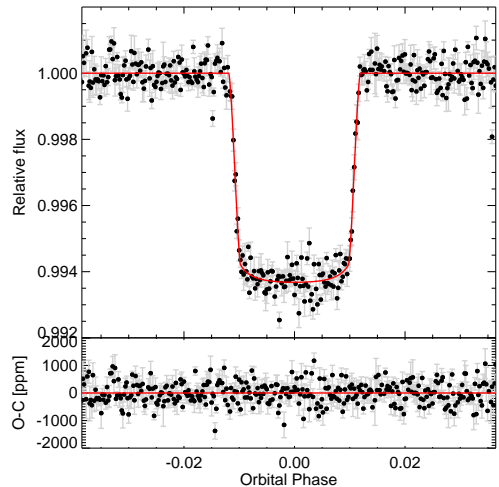}
\figcaption[]{GJ\,3470\,b detrended and phase-folded transit light curve combining our two 4.5\,$\mu$m {\it Spitzer}/IRAC visits, with the best-fit transit model superimposed (see Sect.~\ref{planet}). Data points are binned in 2-minute intervals, and residuals are shown in the bottom panel. \label{fig:phot2}}
\end{figure}

We perform a combined MCMC fit including our two {\it Spitzer} transits and the 61 HARPS radial velocities (RV) published in the discovery paper \citep{Bonfils:2012a}. The main goal of this step is to derive the stellar density from the {\it Spitzer} photometry \citep{Seager:2003}, to enable the derivation of the stellar and planetary physical parameters. The following system parameters (`jump parameters') are left free in the MCMC fit, using uniform priors: the orbital period $P$, transit depth $dF$ (planet-to-star area ratio, $(R_p/R_{\star})^2$), transit duration $W$, time of minimum light $T_0$, impact parameter $b=a \cos i/R_{\star}$, the parameter $K' = K\sqrt{1-e^2}P^{1/3}$, where $K$ is the radial-velocity semi-amplitude, $\sqrt{e} \cos \omega$ and $\sqrt{e} \sin \omega$. We use a quadratic law for the limb-darkening. We draw the theoretical values and corresponding uncertainties of the coefficients $u_1$ and $u_2$ from the tables of \citet{Claret:2011} for the $T_{\rm eff}$, $\log g$, and [Fe/H] determinations reported in Section~\ref{stellar}. We use the resulting distributions for $u_1$ and $u_2$ as normal priors in our MCMC fit. We use the linear combinations $c_1=2u_1+u_2$ and $c_2=u_1-2u_2$ as jump parameters, rather than $u_1$ and $u_2$, to minimize the correlations of the resulting uncertainties \citep{Holman:2006}. At each step of the MCMC fit, the stellar density is derived from this set of parameters and Kepler's third law. We run two chains of $10^5$ steps each, where the first 20\% are discarded. We assess the good convergence and mixing of the chains employing the Gelman-Rubin statistic \citep{Gelman:1992}. We add a 2.0\,m\,s$^{-1}$ jitter contribution in quadrature to the radial-velocity error bars to match the RMS of the residuals. This first combined run yields an eccentricity signal compatible with a circular orbit ($\sqrt{e} \cos \omega = -0.09\pm0.14$ and $\sqrt{e} \sin \omega = 0.00\pm0.22$). We therefore repeat the fit setting $\sqrt{e} \cos \omega$ and $\sqrt{e} \sin \omega$ to zero. The difference in BIC between an eccentric and a circular orbit is $\Delta \rm{BIC} = 15$, translating to an odds ratio of $\sim$1800, hence favoring the circular model we adopt in the following. Our analysis assuming a circular orbit yields a stellar density $\rho_{\star}=2.91^{+0.37}_{-0.33} \rho_{\odot}$ that we use as a constraint for the derivation of the stellar parameters in Sect.~\ref{stellar}. The phase-folded {\it Spitzer} light curve is shown in Figure~\ref{fig:phot2}.

\subsection{Spectroscopic measurements}

\subsubsection{FIRE Spectral Analysis}
\label{fire_analysis}

We measure a number of spectral features in our FIRE spectrum of GJ\,3470 for the purpose of deriving its spectroscopic properties, particularly [Fe/H] and $T_{\rm eff}$, using recent calibrations presented for M dwarfs. We follow closely the prescriptions of \citet{Rojas-Ayala:2012}, \citet{Terrien:2012}, and \citet{Mann:2012a} for measuring equivalent widths (EWs) as well as water indices, and to estimate the pseudo-continuum needed for the above metallicity calibrations. 

For the \citet{Rojas-Ayala:2012} calibration, we measure the Na doublet (2.206 and 2.209\,$\mu$m) and the Ca triplet (2.261, 2.263, and 2.265\,$\mu$m) following the integration limits and continuum points of their Table~2. The pseudo-continuum flux of each feature is taken from a linear fit to the median flux within a 3\,nm region around each continuum point. The water index, H$_{2}$O-K2, is measured following Eq.~5 of \citet{Rojas-Ayala:2012}. For the application of the $H$-band and $K$-band metallicity calibrations of \citet{Terrien:2012}, we measure the EW of the Na (2.2074\,$\mu$m), Ca (1.6159, 1.6203, 2.2638\,$\mu$m), and K (1.5171\,$\mu$m) features following the prescription detailed in Sect.~3.1 of their study. The pseudo-continuum is estimated by fitting a fourth order Legendre polynomial to the regions shown in their Fig.~1(A) for the $H$ band, and in their Fig.~1(B) for the $K$ band. The water indices, H$_{2}$O-H and H$_{2}$O-K, are also measured following the definitions in their paper. For the \citet{Mann:2012a} calibration, the EW of the metal-sensitive features F19 (2.2079\,$\mu$m), F20 (2.3242\,$\mu$m), and F22 (2.3844\,$\mu$m) in the $K$-band are measured using the parameters listed in their Table~5. The pseudo-continuum is measured by a linear fit in the spectral regions specified in their Table~4, immediately redward and blueward of each feature. The water index used for this $K$-band calibration is the same as the one described by \citet{Rojas-Ayala:2012}.

\begin{deluxetable}{lcc}
\tabletypesize{\scriptsize}
\tablecaption{Metallicity Estimates [dex] for GJ\,3470 from Near-Infrared and Visible Spectroscopy.\label{metallicities}}
\tablehead{\colhead{Calibration Reference} & \colhead{Bandpass} & \colhead{Value}}
\startdata
\citet{Rojas-Ayala:2012}	&$K$	&$+0.15\pm0.17$   \\
\citet{Terrien:2012}		&$H$	&$+0.25\pm0.12$   \\	
\citet{Terrien:2012}		&$K$	&$+0.19\pm0.12$   \\
\citet{Mann:2012a}		&$K$	&$+0.32\pm0.11$   \\
\citet{Neves:2012a}		& Visible	&$+0.08\pm0.10$  \\ [+1.5ex]
Adopted		&	\nodata  &$+0.20\pm0.10$ 
\enddata
\end{deluxetable} 
 
The equivalent widths and spectral indices computed from our FIRE spectrum as described above yield the metallicities for GJ\,3470 shown in the first four lines of Table~\ref{metallicities}. A spectroscopic estimate of the effective temperature of GJ\,3470 is obtained using the temperature-sensitive H$_2$O-K2 index in the $K$ band as defined by \citet{Rojas-Ayala:2012}. The result is $3750 \pm 300$\,K.

\subsubsection{HARPS Spectral Analysis}

An additional spectroscopic estimate of the metallicity of GJ\,3470 is obtained from a recent calibration (Neves et al., in prep.) based on the visible-light HARPS spectra of \citet{Bonfils:2012a}. This calibration was established on the basis of equivalent widths measured for a total of 4441 lines in the spectra of 55 stars from the HARPS volume-limited M-dwarf sample \citep{Bonfils:2013}, and is anchored on existing photometric calibrations for metallicity \citep{Neves:2012} and effective temperature \citep{Casagrande:2008a}. The procedure, described briefly in the Appendix of the study by \citet{Neves:2012}, achieves an improved precision over previous methods of 0.10 dex. The result of this measurement for GJ\,3470 is $+0.08 \pm 0.10$, and is collected also in Table~\ref{metallicities}.

The five estimates of [Fe/H] from the FIRE and HARPS spectra are consistent with each other, and we therefore adopt for the remainder of the paper the weighted average, ${\rm [Fe/H]} = +0.20 \pm 0.10$, in which the uncertainty is a more conservative estimate than the formal error of the mean.

\section{Stellar Characterization}
\label{stellar}

Mass ($M_{\star}$) and radius ($R_{\star}$) estimates for exoplanet
host stars are typically obtained by appealing to stellar evolution
models. For M dwarfs this, too, has been problematic (beyond the
challenges for determining $T_{\rm eff}$ and [Fe/H] alluded to earlier
in Sect.~\ref{fire}) because of known disagreements between
predictions from theory and accurate measurements of $M_{\star}$ and
$R_{\star}$ for low-mass stars in double-lined eclipsing binaries
\citep[see, e.g.,][and references therein]{Torres:2012a}. We therefore
rely here exclusively on empirical relations, on the mean stellar
density inferred from our {\it Spitzer\/} light curve in
Sect.~\ref{density} ($\rho_{\star} =
2.91_{-0.33}^{+0.37}\,\rho_{\odot}$), and on brightness measurements
for GJ\,3470 from 2MASS and in the optical \citep[$V = 12.33 \pm
0.01$;][]{Weis:1986, Evans:2002,Zacharias:2013}. For a given parallax
and ignoring extinction, the near-infrared mass-luminosity (M-L)
relations of \cite{Delfosse:2000} provide estimates of the absolute
mass, and are insensitive to metallicity. On the other hand, the
surface-brightness (SB) relations by \cite{Kervella:2004a} allow one
to estimate the angular diameter, which may be converted to a linear
radius with knowledge of the parallax.  The latter relations are valid
for [Fe/H] between $-0.5$ and $+0.5$, and are thus applicable to
GJ\,3470, with its metallicity of ${\rm [Fe/H]} = +0.20 \pm 0.10$.
While a trigonometric parallax has not been measured for this star, we
may use the above relations simultaneously to solve for the distance
that yields values of $M_{\star}$ and $R_{\star}$ consistent with the
measured mean density.

We proceeded in a Monte Carlo fashion, drawing all measured quantities
($VJHK_s$ photometry and {\it Spitzer} mean density) from appropriate
Gaussian distributions. For each set of draws we solve for the value
of the parallax that gives a mass and radius through the M-L and SB
relations resulting in a mean density equal to the randomly drawn
value of $\rho_{\star}$ for the set. We repeat the process $10^5$
times, and adopt as final values the mode of the corresponding
posterior probability distributions, assigning 1-$\sigma$
uncertainties given by the 15.85 and 84.13 percentiles of those
distributions.  We obtain $M_{\star} =
0.539_{-0.043}^{+0.047}\,M_{\odot}$ and $R_{\star} =
0.568_{-0.031}^{+0.037}\,R_{\odot}$, and a parallax of $\pi =
32.4_{-1.9}^{+2.1}$~mas, corresponding to a distance of
$30.7_{-1.7}^{+2.1}$~pc.  The mass is an average of the $J$-, $H$-,
and $K$-band relations by \cite{Delfosse:2000}, each of which is
assumed conservatively to carry an uncertainty of 10\%. The radius is
an average of the two surface-brightness relations of
\cite{Kervella:2004a} that yield the smallest scatter in the angular
diameters (about 1\% for the relations that depend on $V\!-\!H$ and
$V\!-\!K$). Prior to using them, the 2MASS magnitudes are converted
to the native photometric system of the M-L and SB relations (CIT and
Johnson, respectively) using the transformations of
\cite{Carpenter:2001}.  The uncertainties listed above include all
photometric errors, the error in $\rho_{\star}$, as well as the
scatter of the empirical relations. We note that our stellar mass is
very close to that reported by \citet{Bonfils:2012a}, but our radius
is 13\% larger. 

\begin{figure}
\epsscale{1.1}
\plotone{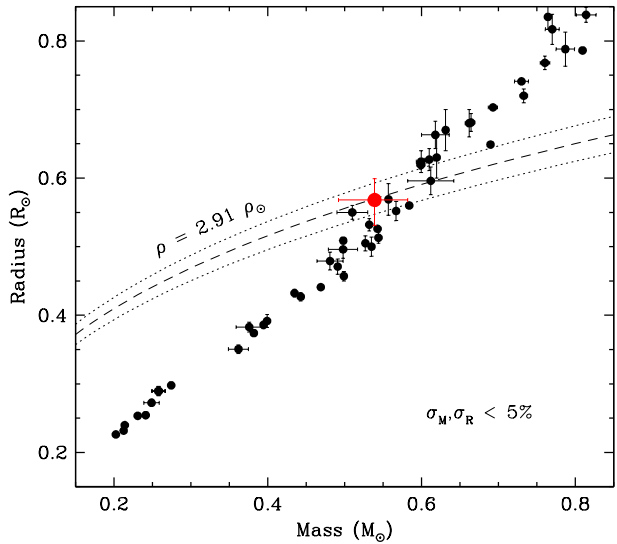}
\figcaption[]{Mass and radius estimates for GJ\,3470 (large red dot)
compared with measurements for other low-mass stars in double-lined
eclipsing binaries with relative errors in $M_{\star}$ and $R_{\star}$
less than 5\% \citep[see][]{Torres:2012a}. The constraint on the mean
stellar density derived from our {\it Spitzer\/} observations is
indicated with the dashed line (dotted lines
representing the 1-$\sigma$ uncertainties).\label{fig:MRstars}}
\end{figure}

As a check on the above absolute radius determination, we obtain
additional estimates of $R_{\star}$ from color indices and the
calibrations recently published by \cite{Boyajian:2012b}, which are
based on angular diameter measurements from the CHARA interferometer
and HIPPARCOS parallaxes, and have a dependence on metallicity.
Results using $V\!-\!J$, $V\!-\!H$, and $V\!-\!K_s$ for the measured
metallicity of GJ\,3470 give very consistent values for $R_{\star}$
averaging $0.513 \pm 0.043\,R_{\odot}$, in agreement with our {\it
Spitzer}-based determination within about 1$\sigma$.
Figure~\ref{fig:MRstars} displays the location of GJ\,3470 in the
mass-radius diagram for low-mass stars, along with the measurements
for all other such objects in double-lined eclipsing binaries that
have relative measurement precisions under 5\% for $M_{\star}$ and
$R_{\star}$.  The constraint afforded by the mean stellar density is
also indicated.

While an estimate of the effective temperature of the star was
obtained earlier using our FIRE spectrum, the precision is relatively
low.  We obtain a further estimate using the color/temperature
calibrations of \citet{Boyajian:2012b}, which are based on bolometric
fluxes and angular diameter measurements, and include metallicity
terms.  The $V\!-\!J$, $V\!-\!H$, and $V\!-\!K_s$ indices along with
our adopted value of [Fe/H] lead to a weighted average temperature of
$3630 \pm 100$\,K.  A final $T_{\rm eff}$ estimate is inferred from
the same three indices and the color/temperature calibrations of
\cite{Casagrande:2008a}, which rely on the Infrared Flux
Method. However, these relations do not take into account the
metallicity, and implicitly assume a composition near solar whereas
GJ\,3470 is metal-rich. Therefore, the resulting estimate ($3360 \pm
100$\,K) requires an adjustment for metallicity.  We determine this
by using the stellar evolution models of \cite{Dotter:2008} in a
differential sense, first reading off from a ${\rm [Fe/H]} = +0.20$
isochrone the stellar mass that yields the same color indices as we
measure, and then comparing the corresponding temperature with that
for a star of the same mass on a solar-metallicity isochrone. This
exercise is insensitive to the age adopted for the
isochrone. Consistent results using the three color indices separately
give an average correction of $+140$\,K, which results in a final
temperature of $3500 \pm 150$\,K. As the two photometric
determinations above are consistent with each other and with the
spectroscopic determination in Sect.~\ref{fire_analysis}, we adopt the
weighted average of the three values, $T_{\rm eff} = 3600 \pm 100$\,K.

While this paper was under review, we learned that \citet{Pineda:2013} 
performed an independent characterization of GJ\,3470's stellar properties.
We refer the reader to that study for a description of their analysis and results.

\section{Planetary and Orbital Parameters}
\label{planet}

Our final parameters for GJ\,3470\,b were derived using the stellar properties from the preceding section, and a MCMC fit analogous to that described in Sect.~\ref{analysis} with the addition of a prior on the stellar mass. For this prior we used the posterior probability distribution derived in Sect.~\ref{stellar}, drawing from it a random value of the mass at each step of the MCMC fit. As before, we included the light curves from both {\it Spitzer} visits, along with the 61 HARPS radial velocities reported by \cite{Bonfils:2012a}. The results are presented in Table~\ref{sysparam}, where the value adopted for each parameter is the median of the corresponding marginalized posterior distribution from the MCMC fit. Error bars are the corresponding 68.3\% probability intervals from the same distributions. The final model and phase-folded {\it Spitzer} light curves are displayed in Figure~\ref{fig:phot2}.

We find for GJ\,3470\,b a radius of $R_{p} =4.83_{-0.21}^{+0.22}\,R_{\oplus}$, which is 13\% larger than previously reported in the literature. This increase is driven mainly by the larger stellar radius from Sect.~\ref{stellar}. Combining the planetary radius with the mass $M_{p} = 13.9^{+1.5}_{-1.4}\,M_{\oplus}$ that relies on the radial-velocity data set yields a very low planetary density of $\rho_{p}= 0.72^{+0.13}_{-0.12}$ g\,cm$^{-3}$, which is 33\% smaller than the estimate in the discovery paper. These planetary parameters are also listed in Table~\ref{sysparam}.

Finally, we performed a new fit for the purpose of assessing the robustness of the orbital period determination for GJ\,3470\,b, which in our solution is constrained both by the two {\it Spitzer} transits and the radial velocities. However, the two {\it Spitzer} visits are consecutive (2012 June 11 and 15), so the lever arm for the orbital period determination is very short. We therefore incorporated the two {\tt TRAPPIST} transit light curves from \citet{Bonfils:2012a}, as well as the ones from {\it EulerCam} and the {\tt NITES} telescope. The light curves from the first two sources show only the ingress portion of the transit, but may still be combined with our two full {\it Spitzer} light curves that constrain the transit shape, if we assume the latter does not change across wavelengths. The {\tt NITES} light curve has a higher level of correlated noise, but does cover the transit completely.
These ground-based light curves were obtained between February and April of 2012, and therefore contribute to build up a much longer baseline.

As expected, most of the system parameters in this new fit are tightly constrained by the {\it Spitzer} photometry alone, but the period is considerably improved. The new value is included in Table~\ref{sysparam}, and is only $19 \pm 11$\,s shorter than the one that relies on the two {\it Spitzer} transits alone. 

\begin{deluxetable}{lc}
\tabletypesize{\scriptsize}
\tablecaption{Adopted system parameters for GJ\,3470 from our MCMC fit of Sect.~\ref{planet}.\label{sysparam}}
\tablehead{\colhead{Parameter} & \colhead{Value} }
\startdata
\textit{Jump parameters} &   \\ 
\tableline
 &   \\ [-1.5ex]
Planet/star area ratio $R_p/R_s$ & $0.07798^{+0.00046}_{-0.00045}$ \\
$b=a \cos i /R_{\star}$ [$R_{\star}$] & $0.40^{+0.06}_{-0.08}$ \\
Transit width $W$ [d] & $0.0791\pm 0.0005$ \\
$T_0 - 2,\!450,\!000$ [BJD$_{\rm TDB}$]& $6090.47701\pm 0.00010$ \\
Orbital period $P$ [d]\tablenotemark{a} & $3.33665\pm 0.00005$  \\ RV $K'$ [m\,s$^{-1}$\,d$^{1/3}$]  & $13.4\pm1.2$  \\
$\sqrt{e} \cos \omega$ & 0.0 (fixed)  \\
$\sqrt{e} \sin \omega$ & 0.0 (fixed)  \\
$c_1 = 2 u_1 + u_2$ & $0.246\pm0.027$  \\
$c_2 = u_1 - 2 u_2$ & $-0.329\pm0.020$  \\
 &    \\
\textit{Stellar parameters} &    \\
\tableline
 &    \\ [-1.5ex]
$u_1$ & $0.033\pm0.015$  \\
$u_2$ & $0.181\pm 0.010$  \\
Mean density $\rho_{\star}$ [$\rho_{\odot}$] & $2.91^{+0.37}_{-0.33}$  \\
Surface gravity $\log g_{\star}$ [cgs] & $4.658 \pm 0.035$  \\
Mass $M_{\star}$ [$M_{\odot}$]\tablenotemark{b}& $0.539^{+0.047}_{-0.043}$  \\
Radius $R_{\star}$ [$R_{\odot}$]\tablenotemark{b}& $0.568^{+0.037}_{-0.031}$  \\
Parallax $\pi$ [mas]\tablenotemark{b}& $32.4_{-1.9}^{+2.1}$ \\
Distance [pc]\tablenotemark{b} & $30.7_{-1.7}^{+2.1}$ \\
Effective temperature $T_{\rm eff}$ [K]\tablenotemark{b}& $3600 \pm 100$\\
Metallicity ${\rm [Fe/H]}$ [dex]\tablenotemark{b} & $+0.20\pm0.10$ \\
 &    \\
\textit{Planetary parameters}  &  \\
\tableline
 &  \\ [-1.5ex]
RV semi-amplitude $K$  [m\,s$^{-1}$]  & $8.9 \pm 1.1$ \\
Orbital semi-major axis $a$ [AU]  & $0.03557^{+0.00096}_{-0.00100}$ \\
Orbital inclination $i$ [deg]  & $88.3^{+0.5}_{-0.4}$ \\
Mean density $\rho_{p}$ [g\, cm$^{-3}$]  &$0.72^{+0.13}_{-0.12}$ \\
Surface gravity $\log g_p$ [cgs]  & $2.76^{+0.06}_{-0.07}$ \\
Mass $M_{p}$ [$M_{\oplus}$]  & $13.9^{+1.5}_{-1.4}$ \\
Radius $R_{p}$ [$R_{\oplus}$]  & $4.83^{+0.22}_{-0.21}$ \\
 &    \\
\textit{Individual transit timings}  &  \\
\tableline
 &  \\ [-1.5ex]
 $T_{0,1} - 2,\!450,\!000$ [BJD$_{\rm TDB}$]& $6090.47705\pm 0.00014$ \\
 $T_{0,2} - 2,\!450,\!000$ [BJD$_{\rm TDB}$]& $6093.81372\pm 0.00015$ 
\enddata
\tablenotetext{a}{Derived using our two {\it Spitzer} light curves along with published ground based photometry and RVs (see Sect.~\ref{planet}).}
\tablenotetext{b}{Parameters derived either in Sect.~\ref{analysis} or in Sect.~\ref{stellar}, and repeated here for convenience.}
\end{deluxetable}

\section{Interior Composition of a Low-density Exo-Neptune}
\label{internal}

\begin{figure}
\epsscale{1.0}
\plotone{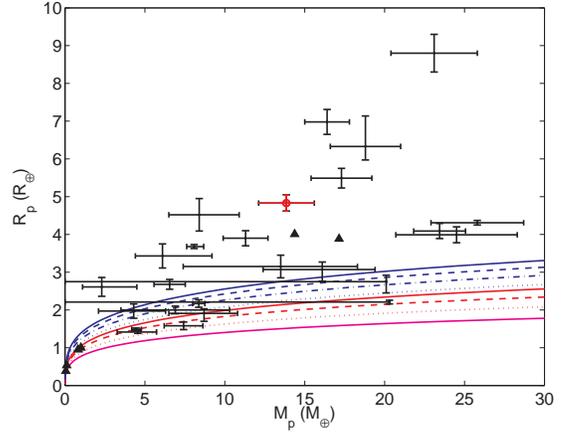}
\figcaption[]{Mass-radius relationships of small transiting planets. GJ\,3470\,b is highlighted in red. Other small transiting exoplanets with dynamical mass constraints (CoRoT-7\,b, Kepler-4\,b, Kepler-10\,b, Kepler-11\,b, c, d, e, f, Kepler-18\,b, c, d, Kepler-19\,b, Kepler-20\,b, c, d, Kepler-30\,b, d, Kepler-36\,b, c, 55\,Cnc\,e, GJ\,1214\,b,  GJ\,436\,b, HAT-P-11\,b, and HAT-P-26\,b) are plotted in black. The solar system planets are indicated with solid triangles.
The curves are illustrative constant-temperature mass-radius relations from \citet{Seager:2007}. The solid lines are homogeneous-composition planets: water ice (blue solid), MgSiO$_3$ perovskite (red solid), and iron (magenta solid). The non-solid lines are mass-radius relations for differentiated planets: 75\% water ice, 22\% silicate shell, and 3\% iron core (blue dashed); Ganymede-like with 45\% water ice, 48.5\% silicate shell, and 6.5\% iron core (blue dot-dashed); 25\% water ice, 52.5\% silicate shell, and 22.5\% iron core (blue dotted); Earth-like with 67.5\% silicate mantle and 32.5\% iron core (red dashed); and Mercury-like with 30\% silicate mantle and 70\% iron core. \label{fig:MR}}
\end{figure}

GJ\,3470\,b presents a valuable test case for planet formation and evolution theories. It stands out from the crowd of accumulating transiting exo-Neptunes due to its low mean density and bright M dwarf host star. 
GJ\,3470\,b's measured radius is $20\%\pm6\%$ larger than Uranus ($R_{\uranus}=4.01~R_{\oplus}$) despite having a similar mass ($M_{\uranus}=14.5~M_{\oplus}$). The planet radius corresponds to roughly 20\% of its Roche-lobe radius. 
Among currently known low-mass ($M_p<30~M_{\oplus}$) transiting planets, only the Kepler-11,  Kepler-18, Kepler-30, and HAT-P-26 systems have planets with lower densities (see Figure~\ref{fig:MR}). 

GJ\,3470\,b must have acquired H/He gas from the protoplanetary nebula. 
 Alternative gas layer sources such as sublimated ices and outgassing from a rocky interior may be important for less massive, more dense planets \citep[such as GJ\,1214b;][]{Rogers:2010a}, but cannot be the dominant gas layer source for GJ\,3470\,b.
Its bulk density is too low for astrophysical ices (H$_2$O, CO$_2$, etc.) alone to comprise the planet volatiles; significant quantities of light gases (hydrogen and helium) must be present. Further, GJ\,3470\,b's gas layer is too voluminous to have been formed by outgassing of light gases during formation; the planet radius exceeds the upper limit for outgassed planets from \citet{Rogers:2011}. 
 
\begin{figure}
\epsscale{1.0}
\plotone{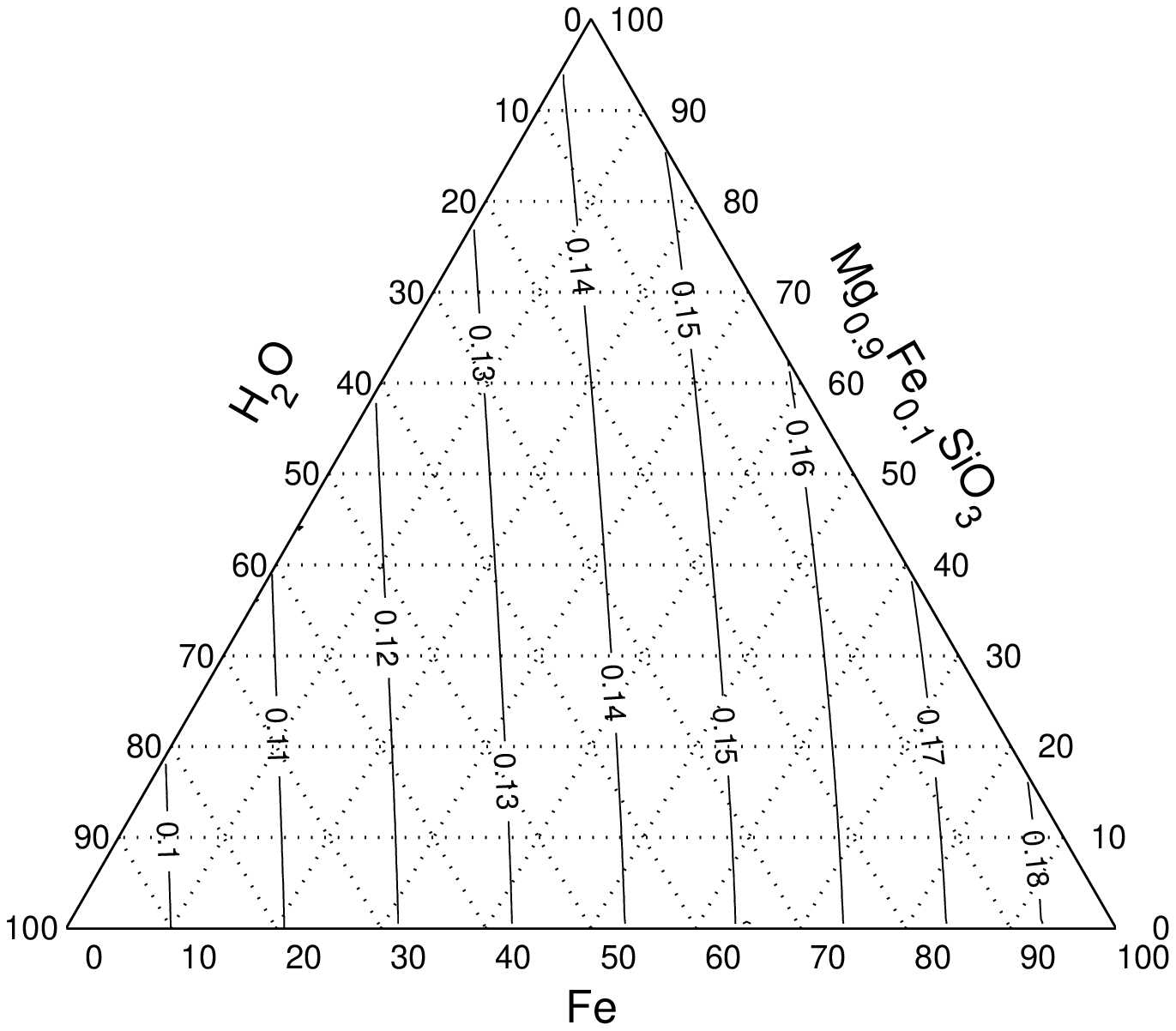}
\figcaption[]{Fraction of GJ\,3470\,b's mass contributed by H/He, as a function of the planet's heavy-element interior composition. Each point within the diagram corresponds to a specific combination of Fe, Mg$_{0.9}$Fe$_{0.1}$SiO$_3$, and H$_2$O (by mass) in the heavy element interior of GJ\,3470\,b. 
For instructions on how to read ternary diagrams see, e.g., \citet{Valencia:2007,Zeng:2008a}. 
 Each contour is labeled with $M_{\rm XY}/M_p$ for our fiducial model parameters (median $M_p$, median $R_p$, $A=0.3$, and $L_p/Mp=10^{-10}~\rm{W\,kg^{-1}}$). Uncertainties in the planet mass, radius, and energy budget can affect  $M_{\rm XY}/M_p$ by $\pm 0.03$ to 0.05.}
\label{fig:HHetern}
\end{figure}

Nebular H/He contributes between 5\% and 24\% to GJ\,3470\,b's mass, according to our interior structure models.
Following \citet{Rogers:2010a}, we apply a fully differentiated model for the planet's interior structure consisting of (from the center of the planet outward) an iron core, silicate layer, ice layer, and H/He gas envelope to explore which bulk compositions are consistent with the measured mass and radius of GJ\,3470\,b. Both the planet's bond albedo $A$ (which scales the equilibrium temperature $T_{\rm eq}=\left(1-A\right)^{1/4}\left(683\pm27\right)~\rm{K}$), and the planet's intrinsic luminosity $L_p$ (a proxy for the poorly constrained age of the planet) are unknown.
We adopt fiducial values of $A=0.3$ and $L_p/M_p=10^{-10}~\rm{W\,kg^{-1}}$, while also exploring the ranges of $A=0$ to 0.5 and $L_p/M_p=10^{-10.5}~\rm{W\,kg^{-1}}$ to $10^{-9.5}~\rm{W\,kg^{-1}}$. Figure~\ref{fig:HHetern} presents the H/He gas mass fraction ($M_{\rm XY}/M_p$) in our models as a function of the Fe-silicate-H$_2$O abundances of the heavy element interior (assuming the median values of the planet mass and radius, and our nominal planet energy budget parameters). Varying the planet mass and radius within their 1-$\sigma$ bounds, and considering a range of plausible planet energy budgets affects the H/He mass fractions by up to $\pm0.05$. For a rocky Earth-like heavy element interior composition (32\% Fe, 68\% silicate, 0\% H$_2$O), GJ\,3470\,b's H/He envelope mass is constrained to $M_{\rm XY}/M_p = 0.16\pm0.05$, while for a denser iron-enhanced Mercury-like rocky interior  (70\% Fe, 30\% silicate, 0\% H$_2$O),  $M_{\rm XY}/M_p = 0.17\pm0.05$. Less H/He is needed if GJ\,3470\,b has an ice-rich interior composition; for instance, for a heavy element interior with 16\% Fe, 34\% silicate, 50\% H$_2$O, $M_{\rm XY}/M_p = 0.12^{+0.05}_{-0.04}$. 

Which heavy element interior compositions are plausible for GJ\,3470\,b? The planet interior ice-to-rock ratio is not constrained by measurements of the planets mass and radius alone, so we look to planet formation theory for insights. If GJ\,3470\,b formed beyond the snow line and migrated inwards to its current orbit, its heavy element interior would be ice-rich. 
If instead GJ\,3470\,b formed in situ (inside the snow line) its heavy element interior would be rock dominated with a lower proportion of ices. Theoretical predictions for how much ice is likely included in planets formed inside the snow line of M dwarfs are a topic of ongoing debate. \citet{Ogihara:2008} proposed that migration of planetesimals from beyond the snow line could supply icy material to the inner regions of the protoplanetary disk. On the other hand, \citet{Lissauer:2007} and \citet{Kennedy:2007} predict that planets and planetesimals formed within 1\,AU of M dwarfs are unlikely to have large volatile inventories when the effect of the M dwarfs' pre-main sequence luminosity evolution is taken into account. In Figure~\ref{fig:HHetern} we present interior bulk compositions for the full range of ice-to-rock ratios.

\section{Summary}

Our 4.5\,$\mu$m {\it Spitzer} observations have enabled us to refine the planetary and system parameters of the Neptune-size planet GJ\,3470\,b, improving its radius to $R_{p} = 4.8\pm0.2 R_{\oplus}$, which is 13\% larger than previously reported in the literature. As a result, the revised planetary density, $\rho_{p}= 0.72\pm0.13$ g\,cm$^{-3}$, is 33\% smaller than before. These changes come mostly from revisions of the stellar parameters (particularly $R_{\star}$), which have been frustratingly difficult to determine accurately in the past due to known discrepancies between observations and standard stellar evolution models for lower main-sequence stars. In this paper we have relied for this only on empirical mass-luminosity and surface brightness relations that have been widely employed in other contexts, and on the strong constraint on the mean stellar density provided by our {\it Spitzer} observations. In the process we have inferred an accurate distance for the star.

GJ\,3470\,b provides a valuable example of an extremely low-density planet, representative of a significant portion of the exoplanet candidates found by the {\it Kepler} mission to date. The brightness of the host star ($K_s = 7.99$) combined with its large planet-to-star radius ratio renders GJ\,3470\,b a promising candidate for future atmospheric characterization, which could provide clues on its formation pathway. Indeed, GJ\,3470\,b's low surface gravity translates to a large atmospheric scale height for a given atmospheric composition, favoring follow-up studies applying transmission spectroscopy. GJ\,3470\,b, GJ\,436\,b, and GJ\,1214\,b are a remarkable sample of volatile-rich planets orbiting bright nearby stars, pushing the field of comparative exoplanetology further towards low-mass planets.

\acknowledgments

We thank Zach Berta and Elisabeth Newton for helpful discussions regarding near-IR M-dwarf spectral characterization. We are grateful to Rob Simcoe, Paul Schechter, Elisabeth Adams and David Ciardi for their help in obtaining the ground-based observations presented in this paper. We thank the anonymous referee for a report that improved the paper. We thank the \textit{Spitzer} Science Center staff, and especially Nancy Silbermann, for the efficient scheduling of our observations. We also wish to thank the staff of the Magellan Telescopes and Las Campanas Observatory for their assistance in obtaining the FIRE observations. This work is based on observations made with the {\it Spitzer} Space Telescope, which is operated by the Jet Propulsion Laboratory, California Institute of Technology, under a contract with NASA. Support for this work was provided by NASA through an award issued by JPL/Caltech. The WIYN Observatory is a joint facility of the University of Wisconsin-Madison, Indiana University, Yale University, and the National Optical Astronomy Observatory. Support for LAR was provided by NASA through Hubble Fellowship grant HF-51313.01-A awarded by the Space Telescope Science Institute, which is operated by the Association of Universities for Research in Astronomy, Inc., for NASA, under contract NAS 5-26555. This work was supported by the European Research Council/European Community under the FP7 through Starting Grant agreement number 239953 and by Funda\c{c}\~ao para a Ci\^encia e a Tecnologia (FCT) in the form of grants PTDC/CTE-AST/098528/2008 and PTDC/CTE-AST/120251/2010. The financial support from the ``Programme National de Plan\'etologie'' (PNP) of CNRS/INSU, France, is gratefully acknowledged. GT acknowledges partial support for this work through NSF grant AST-1007992. VN would also like to acknowledge the support from the FCT in the form of the fellowship SFRH/BD/60688/2009. MG is Research Associate at the Belgian Fonds National de la Recherche Scientifique (FNRS).

{\it Facilities:} \facility{{\it Spitzer}, Magellan, WIYN}

\bibliography{apj-jour,gj3470}

\end{document}